\documentstyle[bezier]{mn}
\input{epsf}

\newcommand{\vB}{\mbox{\boldmath $B$}}
\newcommand{\vV}{\mbox{\boldmath $V$}}

\newcommand{\grad}{\mbox{\boldmath$\nabla$}}
\newcommand{\cross}{\mbox{\boldmath$\times$}}
\newcommand{\curl}{\mbox{\boldmath$\nabla\times$}}
\newcommand{\diver}{\mbox{\boldmath$\nabla\cdot$}}
\newcommand{\pd}{\partial}
\newcommand{\Pm}{P_{\rm m}}
\newcommand{\Rm}{R_{\rm m}}

\begin{document}

\title[MRI in Liquid Metal Annulus]{Magnetorotational Instability in a Rotating Liquid Metal Annulus}
\author[Ji, Goodman, and Kageyama]{Hantao Ji,$^1$\thanks{E-mail: hji@pppl.gov}
Jeremy Goodman$^2$ and Akira Kageyama$^{1,3}$\\
$^1$ Princeton Plasma Physics Laboratory, Princeton, NJ 08543, USA\\
$^2$ Princeton University Observatory, Princeton, NJ 08544, USA\\
$^3$ National Institute for Fusion Science, Toki, Gifu 509-5292, Japan}

\maketitle
\begin{abstract}
Although the magnetorotational instability (MRI) has been widely accepted as a
powerful accretion mechanism in magnetized accretion disks, it has not 
been realized in the laboratory. The possibility of studying MRI in a rotating
liquid-metal annulus (Couette flow) is explored 
by local and global stability analysis.
Stability diagrams are drawn in dimensionless parameters, and also in terms
of the angular velocities at the inner and outer cylinders.
It is shown that MRI can be triggered in a moderately rapidly rotating
table-top apparatus, using easy-to-handle metals such as gallium.
Practical issues of this proposed experiment are discussed. 
\end{abstract}


\section{INTRODUCTION}

Astrophysical magnetic fields have long been recognized to be important
but difficult to understand.
A prominent example concerns accretion disks \cite{SS73,LP74}, 
where orbiting plasmas gradually accrete onto a central mass.
Three types of central object occur: protostars (stars in formation), 
collapsed stars in binary systems (white dwarfs,
neutron stars, and black holes),
and supermassive black holes in active galactic nuclei (quasars \emph{et al.}).
In addition to the accretion process,
jets and other spectacular phenomena
may be magnetically driven by the disk \cite{MK01}.
Understanding the dynamics and dissipation
mechanisms in accretion disks holds an important key to understanding many
active astronomical systems as a whole.

The accretion rates cannot be due to ordinary molecular or plasma
viscosity because of the extraordinarily high Reynolds numbers
involved.  Theorists have often appealed to hydrodynamic turbulence
\cite{PR81}, but recent numerical simulations indicate that
nonmagnetic disks are stabilized by their positive angular momentum
gradient \cite{BH96,Ca96,HB99}; in effect, Rayleigh's
stability criterion appears to suppress local nonaxisymmetric
as well as axisymmetric disturbances \cite{RA16}.
Linear axisymmetric instability of a \emph{magnetized} but Rayleigh-stable 
fluid, the magnetorotational instability (MRI), was discovered decades
ago \cite{VE59,CH60} but did not come to the attention of the
astrophyisical community until recently rediscovered \cite{BH91},
despite general recognition that magnetic effects might somehow be
important \cite{SS73}.  Since then, many analytic and numerical
studies of the MRI have been performed under increasingly complex and
realistic assumptions, including such effects as finite resistivity,
global boundary conditions, and nonlinearity in two and three
dimensions
\cite{BH91b,CP94,BB94,BR95,MT95,HG96,SH96,GA96,JI96,SM99,FS00,HA00}.

Despite its popularity and importance, however, the MRI has never been
realized in the laboratory or demonstrated observationally. Laboratory
plasma experiments are primarily magnetically driven, and the observed
flows, often induced as secondary effects of other instabilities, are
small compared to the Alfv\'en speed. On the other hand, the existing
body of experimental work on magnetized Couette flow using liquid
metals \cite{DO60,DO62,DC64,BR70} has focused on magnetic
stabilization of the Rayleigh instability, as first analyzed by
Chandrasekhar \shortcite{CH61}. In this Letter, we explore the
feasibility of a Couette-flow experiment dedicated to MRI.

\section{LOCAL STABILITY ANALYSIS}

Couette flow involves a liquid confined between rotating
coaxial cylinders \cite{CO90}.
Let their radii be $r_1<r_2$, and their angular velocities
$\Omega_1$, $\Omega_2$.
In steady state, the
radial angular momentum flux, $ h \cdot 2\pi r \cdot \rho\nu \cdot r^2
(-\pd \Omega /\pd r)$, is constant with radius, where
$h$ is the depth of the liquid, $\rho$ is its density, and
$\nu$ is its kinematic viscosity.  If $\pd h/\pd r=0$, then the angular
velocity of the liquid satisfies $r^3 \pd \Omega / \pd r = $const.,
so that
\begin{equation}
\Omega(r)=a+{b \over r^2},
\label{Couette}
\end{equation}
where $a = (\Omega_2 r_2^2 - \Omega_1 r_1^2)/(r_2^2-r_1^2) $ and 
$b = r_1^2r_2^2(\Omega_1-\Omega_2)/(r_2^2-r_1^2).$
The Rayleigh stability criterion is $a\Omega>0$.

The dynamics of liquid metals is well described by the
incompressible and dissipative magnetohydrodynamic (MHD) equations,
\begin{eqnarray*}
0 &=& \diver \vV  \\
0 &=& \diver \vB  \\
{\pd \vB \over \pd t} &=&
\curl (\vV \cross \vB) +\eta\grad^2 \vB  \\
{\pd \vV \over \pd t} + (\vV \cdot \grad) \vV &=&
{(\vB \cdot \grad) \vB \over\mu_0\rho} -{1 \over \rho}\grad 
\left(p+\frac{B^2}{2\mu_0}\right)\\ 
&~& +\nu \grad^2 \vV, 
\end{eqnarray*}
where $\vV$ is velocity, $\vB$ is magnetic field, $\eta$ is magnetic diffusivity,
and $p$ is a scalar potential incorporating both pressure and gravity.
In cylindrical coordinates, the equilibrium quantities are
$\vB_0 = (0,0,B)$ and $\vV_0 = (0, r\Omega, 0)$, and the balance of forces is
$\pd p_0 / \pd z = 0$ and $\pd p_0 / \pd r = \rho r \Omega^2$.

WKBJ methods describe the stability of this system very well even on
the largest scales.  Using cylindrical coordinates, the perturbations
are $\vB_1 = (B_r, B_\theta, B_z)$ and $\vV_1 = (V_r, V_\theta, V_z)$,
all proportional to $\exp(\gamma t\,-ik_z z\,-ik_r r )$, so that
$\gamma$ is the growth rate and the perturbations are axisymmetric.
The minimum $k_z$ and $k_r$ are assumed to be $\pi/h$ and
$\pi/(r_2-r_1)$, respectively, so that the total wavenumber
$k=\sqrt{k_z^2+k_r^2}=k_z\sqrt{1+\epsilon^2}$, where $\epsilon \equiv
h/(r_2-r_1)$ is the elongation of a toroidal cross-section of the
liquid metal annulus.
The linearized equations of motion are
\begin{eqnarray}
0 & = & k_r V_r + k_z V_z \nonumber \\
0 & = & k_r B_r + k_z B_z \nonumber \\
\gamma B_r & = & -i k_z B V_r -\eta k^2 B_r \nonumber \\
\gamma B_\theta & = & -i k_z B V_\theta + {\pd \Omega \over \pd \ln r}B_r
-\eta k^2 B_\theta \nonumber \\
\gamma V_r - 2 \Omega V_\theta & = & -i {k_z B \over \mu_0 \rho} B_r 
+ i {k_r \over \rho} p_1 + i {k_r B \over \mu_0 \rho} B_z
-\nu k^2 V_r \nonumber \\
\gamma V_\theta + {\kappa^2 \over 2 \Omega} V_r & = & 
-i {k_z B \over \mu_0 \rho} B_\theta -\nu k^2 V_\theta \nonumber \\
\gamma V_z & = & i {k_z \over \rho} p_1 -\nu k^2 V_z \nonumber
\end{eqnarray}
where the epicyclic frequency is defined by
$\kappa^2 \equiv (1/r^3) {\pd (r^4\Omega^2)/ \pd r} = 4\Omega^2+
{\pd \Omega^2 / \pd \ln r}$ and $p_1$ is the perturbed pressure.
The vertical induction equation is not needed since $B_z$ can be deduced from
$\diver\vB=0$.
These equations lead to the following dispersion relation:
\begin{eqnarray*}
[(\gamma +\nu k^2)(\gamma +\eta k^2)+ (k_zV_A)^2]^2{k^2 \over
k_z^2} &+& \kappa^2(\gamma +\eta k^2)^2\\
&+& {\pd \Omega^2 \over \pd \ln r}(k_zV_A)^2=0.
\end{eqnarray*}
The Alfv\'en speed is $V_A \equiv B/\sqrt{\mu_0 \rho}$.
This dispersion relation is identical to the one derived for
accretion disks in the incompressible limit \cite{SM99}.

Introducing a dimensionless vorticity parameter,
$\zeta \equiv (1/r\Omega)\pd (r^2\Omega)/\pd r = 2 + \pd \ln \Omega / \pd \ln r$,
we have $\kappa^2=2\Omega^2\zeta$ so that the Rayleigh stability criterion becomes
$\zeta \ge 0$. Similarly, $\pd \Omega^2 / \pd \ln r = 2\Omega^2(\zeta-2)$.
There are three other relevant frequencies:
resistive, $\omega_\eta \equiv \eta k^2$;
viscous, $\omega_\nu \equiv \nu k^2$; and
Alfv\'enic  $\omega_A \equiv |k_z V_A|$.
Because liquid metals are far more resistive than viscous,
$\omega_\eta$ serves as a base frequency in the following
three dimensionless parameters:
magnetic Prandtl number, $\Pm \equiv \omega_\nu/ \omega_\eta$;
Lundquist number, $S \equiv \omega_A /\omega_\eta$; and
magnetic Reynolds number, $\Rm \equiv \Omega/\omega_\eta$.
The astrophysical literature gives several inequivalent definitions
of ``magnetic Reynolds number,'' some corresponding to our $S$.
Some involve the sound speed, $c_s$, for although 
MRI is essentially noncompressive, vertical force balance
in an astrophysical disk relates $c_s$ to the half-thickness: 
$c_s\approx h\Omega$.
The free energy for MRI derives from differential rotation,
represented in our dimensionless system by a combination of
$\zeta$ and $\Rm$; but magnetic field, represented by $S$, is
required to liberate this energy from hydrodynamic constraints.

Using the normalized growth rate, $\gamma/\omega_\eta \rightarrow \gamma$,
the dispersion relation can be rewritten as
\begin{eqnarray*}
\left[(\gamma + \Pm)(\gamma + 1) + S^2\right]^2(1+\epsilon^2)
\qquad&&\nonumber\\
+2\zeta \Rm^2(\gamma+1)^2 -2(2-\zeta)\Rm^2S^2&=&0.
\end{eqnarray*}
It can be shown that there are no purely imaginary roots for $\gamma$ as follows.
Suppose that $\gamma=i\omega$ for real and nonzero $\omega$.
From the imaginary part of the equation above, one finds
\begin{displaymath}
\omega^2=S^2+\Pm+(1+\Pm)\zeta \sigma^2,\qquad
\sigma\equiv {\Rm \over 1+\Pm} \sqrt{2 \over 1+\epsilon^2}.
\end{displaymath}
Substituting this into the real part yields
\begin{displaymath}
\sigma^4 \zeta^2 + 2\sigma^2 \zeta +1+{S^2 \over \Pm} + 2 {S \sigma^2 \over \Pm} =0.
\end{displaymath}
This is a quadratic equation in $\zeta$ which must be real. 
But the discriminant
\begin{displaymath}
-\sigma^4 \left({S^2 \over \Pm} + 2 {S\sigma^2 \over \Pm} \right),
\end{displaymath}
is negative unless $S=0$ or $\Rm=0$.
Hence there are no purely oscillatory modes.
On the other hand, all roots of the dispersion relation for $\gamma$
have negative real parts as $\Rm\to0$ (nonrotating flow).  
Hence the transition to instability occurs through $\gamma=0$, and
the necessary and sufficient condition for stability is that
the value of the dispersion relation at $\gamma=0$ remain
positive as $\Rm$ increases:
\begin{equation}\label{stabcond}
(\Pm + S^2)^2 (1+\epsilon^2) + 2\zeta \Rm^2 - 2(2-\zeta)\Rm^2 S^2 \ge 0,
\end{equation}
which can be taken into various limits.

{\bf Nonmagnetic limit}. As $\eta \rightarrow \infty$,
the three terms $S^4$, $\Rm S^2$, and $\Rm^2S^2$ approach zero faster than the others,
leading to the stability condition
$\Pm^2 (1+\epsilon^2) + 2\zeta \Rm^2 \ge 0$.  Stability occurs when $\zeta\ge 0$,
and also when $\zeta<0$ if
the Taylor number-like expression $-2\zeta\Omega^2 /\nu^2 k^4 \le 1+\epsilon^2$ \cite{TA23}.

{\bf Ideal MHD limit}.
As $\eta \rightarrow 0$, the other two terms dominate, with stability for
$S^2 (1+\epsilon^2) \ge 2(2-\zeta)\Rm^2$.
Instability occurs at sufficiently weak fields (small $S$) unless
$\zeta\ge 2$, \emph{i.e.} $\pd \ln \Omega /\pd \ln r \ge 0$ \cite{BH91}.

{\bf Small $\Pm$ limit.} In liquid metals, usually viscosity is much smaller
than resistivity, $\Pm \sim 10^{-6}$. As $\Pm\to 0$, eq.~(\ref{stabcond}) reduces to
\begin{equation}
\zeta \ge {2S^2 \over S^2+1} - {S^4 (1+\epsilon^2) \over 2 \Rm^2 (S^2+1)}\,.
\label{condition}
\end{equation}

\begin{figure}
\centerline{\epsfxsize=2.7in\epsffile{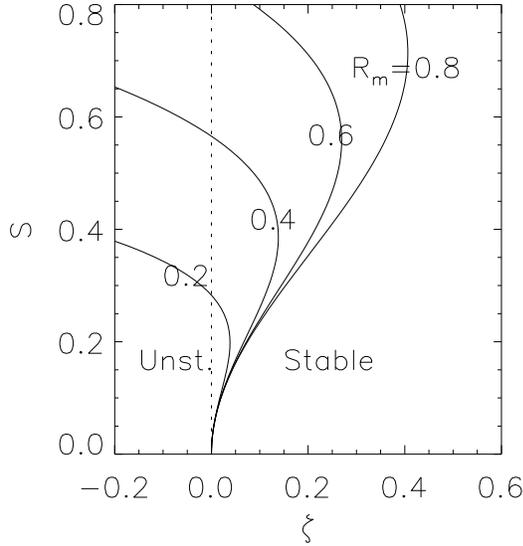}}
\caption{Stability of a rotating liquid metal annulus in dimensionless parameter
space of $(S, \zeta)$ at a few values of $\Rm$ for the case of $\epsilon=1$. 
Areas to right of the curves indicate stability.}
\end{figure}

\section{STABILITY DIAGRAMS AND GROWTH RATES}

The stability condition (\ref{condition}) defines
a two-dimensional surface in the parameter space $(S, \zeta, \Rm)$ at fixed $\epsilon$. 
To illustrate the dependence on these parameters, we vary only two of them at a time.
Stability boundaries in the $(S,\zeta)$ plane are shown in Fig.~1 for the case of $\epsilon=1$.
When $\zeta < 0$, the annulus is unstable hydrodynamically to the Rayleigh mode at $S=0$
but can be stabilized \cite{CH61} by a large magnetic field (large $S$).
When $\zeta > 0$, the annulus is stable at zero field but unstable at some $S>0$ if
$\Rm$ is large enough.
Stability returns at even larger $S$.
The unstable region extends to larger $S$ and $\zeta$ at larger $\Rm$. 
Stability at $S=0$ and as $S \rightarrow \infty$ are hallmarks of MRI \cite{BH91,BH98}.
(In ideal MHD, instability extends formally to $S=0^+$.)
It can be seen that there is a maximum $\zeta$ above which MRI is absent
for a given $\Rm$. From eq.(\ref{condition}),
\begin{equation}
\zeta_{\rm max} = 2- {1+\epsilon^2 \over \Rm^2 } \left( \sqrt{1+{4\Rm^2 \over 1+\epsilon^2}}-1\right)
\label{zetamax}
\end{equation}
at the $S$ value given by $ S^2=\sqrt{1+4\Rm^2/(1+\epsilon^2)}-1$.

\begin{figure}
\centerline{\epsfxsize=2.7in\epsffile{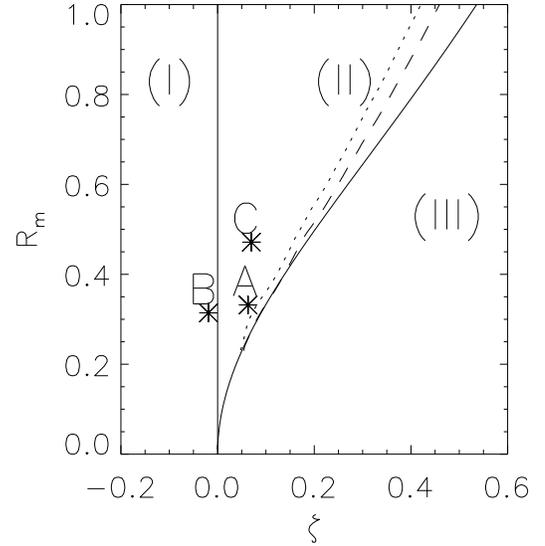}}
\caption{Stability of a rotating liquid metal annulus in $\Rm$ and $\zeta$ space. Here
the stability can be divided into 3 regions: region (I) is hydrodynamically
unstable but can be stabilized by a finite magnetic field, as exemplified by
point B. Region (II) is hydrodynamically stable but can be destabilized by presence of
a magnetic field (MRI), as exemplified by points A and C.
Region (III) is always stable. Results from global eigenmode analysis are also
shown: dotted lines for conducting boundary conditions
and dashed lines for insulating boundary conditions.}
\end{figure}

Figure~2 shows stability in the $(\Rm,\zeta)$ plane.
Region (I) is hydrodynamically unstable but can be stabilized by 
a finite magnetic field. This region has been extensively studied both
theoretically and experimentally \cite{CH61}, and is exemplified by point B. 
Region (II) is hydrodynamically stable but destabilized by
a magnetic field. This is the MRI region, and has never been studied experimentally. 
Growth rates at points A and C are given below.
Region (III) is always stable. The boundary between regions (II) and 
(III) is given by eq.~(\ref{zetamax}).

\begin{figure}
\centerline{\epsfxsize=2.8in\epsffile{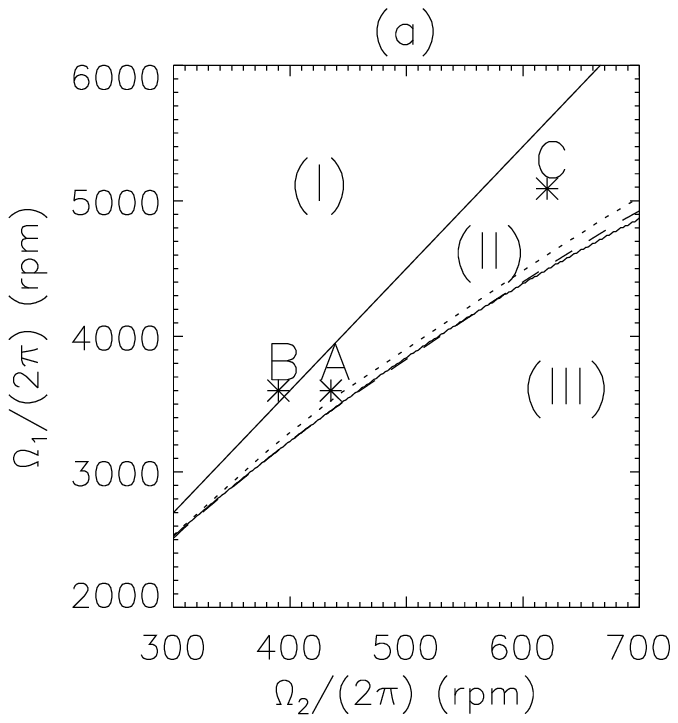}}
\centerline{\epsfxsize=2.6in\epsffile{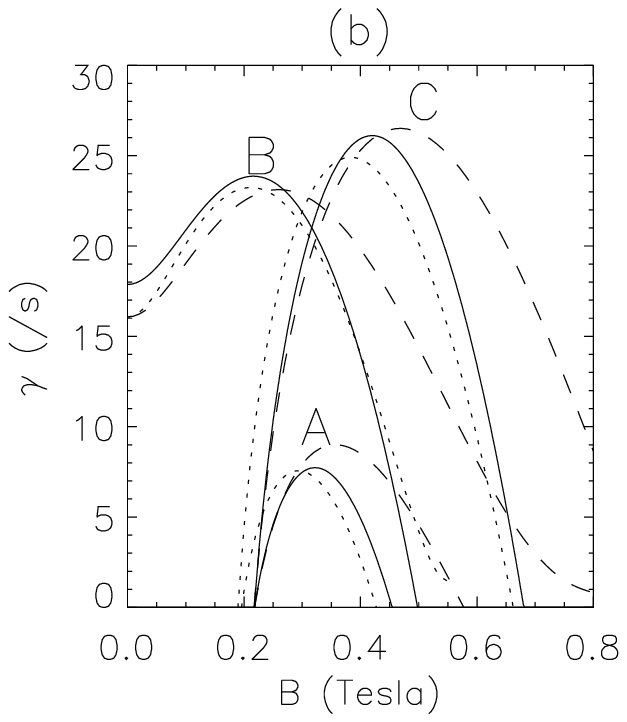}}
\caption{Stability diagram of a rotating gallium annulus in $\Omega_1$ and $\Omega_2$ space
with dimensions $r_1$=0.05m,
$r_2$=0.15m, and $h=$0.1m. The growth rates of points A, B, and C, corresponding to
those in Fig.2, are also shown as functions of magnetic field in (b).
Results from global eigenmode analysis are also
shown: dotted lines for conducting boundary conditions
and dashed lines for insulating boundary conditions.}
\end{figure}

It is useful to project the stability diagram onto experimentally
controllable parameters.  To apply the local dispersion relation, we
take $\zeta = 2a / \bar\Omega$ [from eq.(\ref{Couette})] and $\Rm
\equiv \bar\Omega/\omega_\eta$, with $\bar\Omega\equiv
\sqrt{\Omega_1\Omega_2}$.  Figure~3(a) shows stability in the plane
$(\Omega_1, \Omega_2)$ for an annulus of dimensions $r_1$=0.05m,
$r_2$=0.15m, and $h$=0.1m (hence $\epsilon=1$) filled with gallium
($\rho\simeq 6\times 10^3$kg/m$^3$, $\eta\simeq 0.2$m$^2$/s, $\nu
\simeq 3\times 10^{-7}$m$^2$/s).  Table 1 lists the physical
parameters at points A, B, and C.  The corresponding growth rates are
shown as functions of magnetic field in Fig.3(b).

\begin{table}
\caption{Parameters for a gallium annulus with $r_1=0.05$m, $r_2=0.15$m, and $h=0.1$m.}
\begin{center}
\begin{tabular}{|c|r|r|r|r|}
point & $\Omega_1$(rpm) & $\Omega_2$(rpm) & $\Rm$ & $\zeta$ \\
A & 	3600.00 & 	435.00 & 	0.3319 & 0.06293	\\
B &	3600.00 &	390.00 &	0.3143 & -0.01899	\\
C & 	5089.77 & 	620.70 & 	0.4715 & 0.06984	\\
\end{tabular}
\end{center}
\end{table}

The applicability of WKBJ is subject to doubt, since the most unstable
wavelengths are larger than the gap width and cylinder height.  In
fact, global analysis shows that the eigenfunctions are nonsinusoidal
and sensitive to the boundary conditions. Yet the growth rates are
remarkably robust.  A linearized, finite-difference, initial-value
code was written to to detect the fastest growing mode.  Periodic
boundary conditions were used in $z$.  Radial boundaries were
impenetrable and no-slip ($\delta\vV_1=0$), and electrically either
perfectly insulating or perfectly conducting.  Results are compared
with the WKBJ analysis in Figs.2 and 3.  Figure~4 shows eigenmodes for
the parameters of point C with $B$=0.3 Tesla.  Differences between the
conducting and insulating cases can be seen near the inner
boundaries. A Hartmann layer \cite{HA37}, consisting of large toroidal
and axial velocities within a radial thickness of $\sim \sqrt{\nu
\eta}/V_A <1$mm (not visible in Fig.~4), forms at the inner conducting
boundary as the Lorentz force, $\delta j_r \times B_z$, balances with
the viscous force.  Nevertheless, the growth rates are remarkably
similar to those of the local analysis, which therefore should suffice
for preliminary experimental design.  [Details of the global analysis
will be reported elsewhere \cite{GJ01}.]

\begin{figure}
\centerline{\epsfxsize=3.0in\epsffile{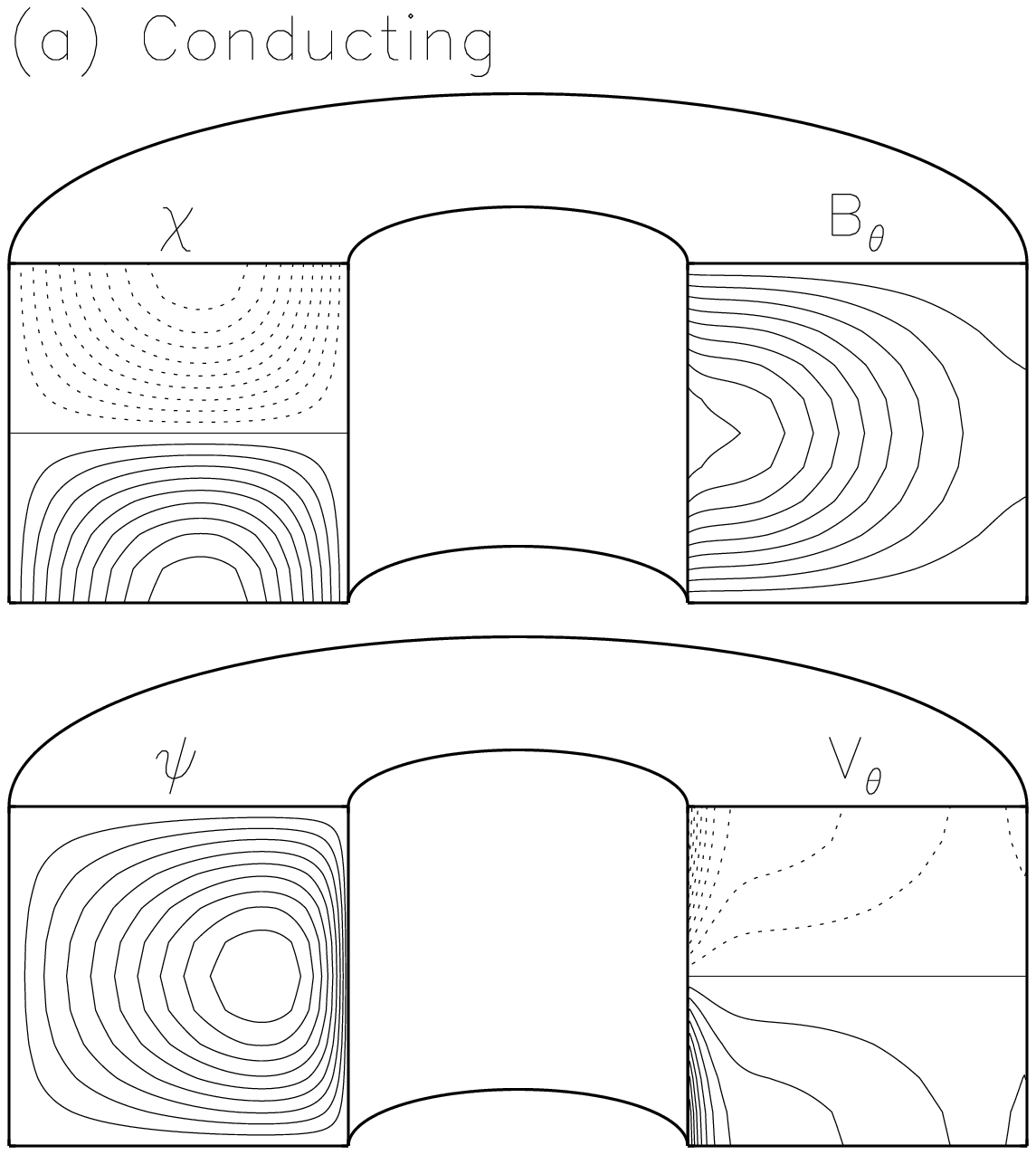}}
\centerline{\epsfxsize=3.0in\epsffile{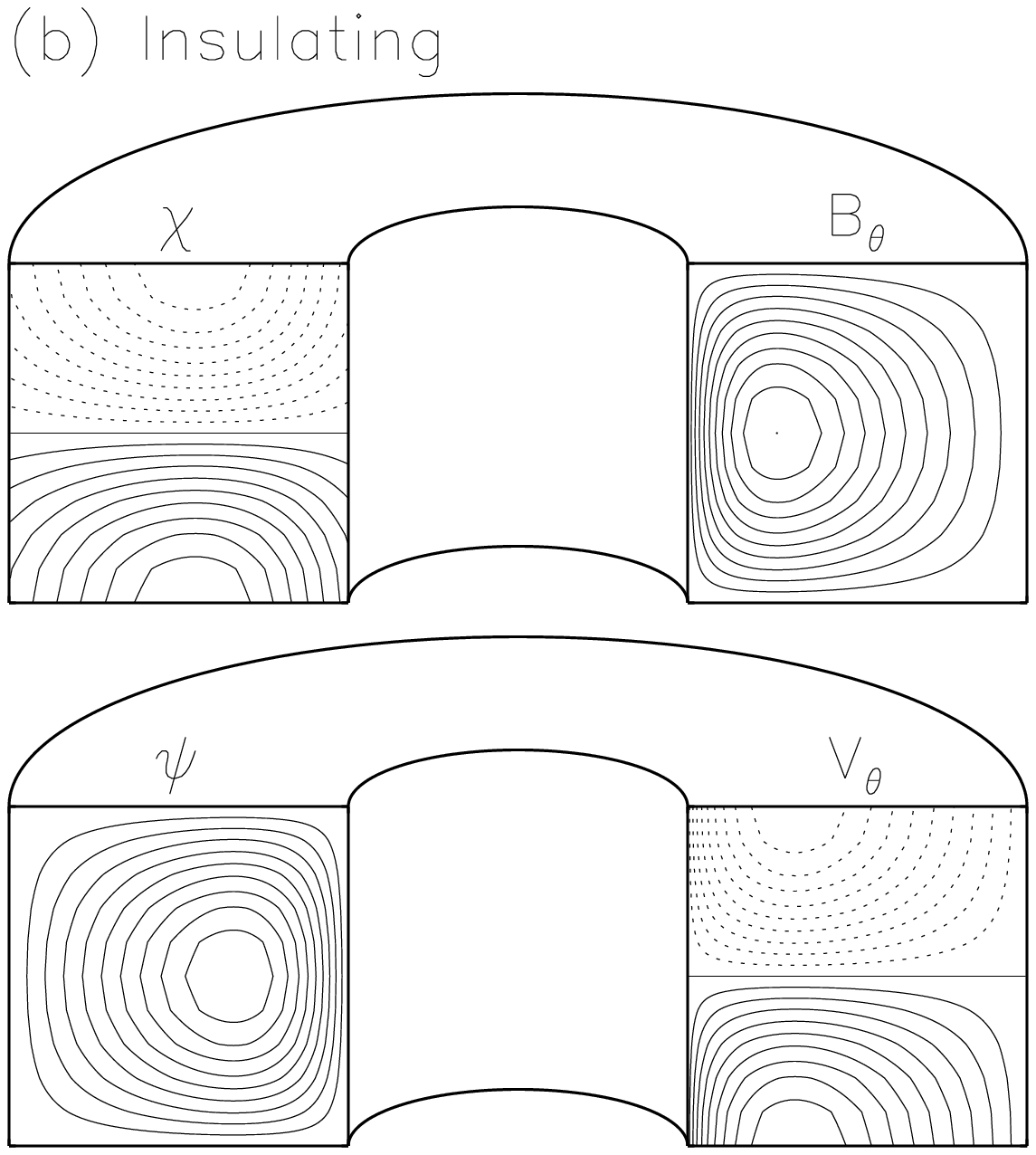}}
\caption{Eigenmodes for conditions given by point C in Fig.3 at $B=0.3$ Tesla with
conducting (a) and insulating (b) radial boundaries. Here, solid (dotted) lines
represent positive (negative) values;
$\chi$ and $\psi$ are poloidal flux and stream functions, respectively.}
\end{figure}

A fully nonlinear incompressible MHD code has been developed
to study the problem in three dimensions \cite{KJ01}. Initial results
from linear and two-dimensional runs of this code with conducting,  freely-slipping
boundary conditions agree with the local and global analyses.
For example, for the conditions given by the
point C in Figs.2 and 3 at $B=0.3$\,Tesla, the growth rate is 21.67 s$^{-1}$ from simulations,
21.90 s$^{-1}$ from global analysis, and 19.10 s$^{-1}$ from local analysis.

\section{DISCUSSIONS AND CONCLUSIONS}

Several issues must be explored before committing to an experimental
design.  The first is geometric optimization with regard to aspect
ratio [$A\equiv (r_2+r_1)/(r_2-r_1)$] and elongation [$\epsilon\equiv
h/(r_2-r_1$)].  Obviously, $\epsilon\sim 1$ is desirable to minimize
volume, and therefore expense, at a given growth rate.  Less
obviously, aspect ratios close to unity cause the eigenmodes and growth rates
to be dominated by the inner cylinder, which is undesirable if the aim
is to imitate relatively uniform local conditions within an
astrophysical disk.  Therefore, moderate values of $A (\sim 2)$ and
$\epsilon (\sim 1)$ are preferred.

The periodic vertical boundary conditions used in all of our analyses
take no account of viscous layers at the top and bottom of the flow.
(The top will have to be capped because of large radial pressure
gradients.) The main effect of these viscous boundary layers is to drive
Ekman circulation, which flows more rapidly against a weak
angular-momentum gradient than against uniform rotation. 
The thickness of the Ekman layer
$\delta_E\approx\zeta^{-1/4}\sqrt{\nu/\bar\Omega}$ is small ($\sim
10^{-3}\,h$ at point C), and the Ekman circulation time
$\zeta^{3/4}h/\sqrt{\nu\bar\Omega}\sim$ 2~s is much longer than a typical
MRI growth time, so we do not expect these layers to be important for
stability.  To further minimize their effect, if necessary, one could
increase $\epsilon$, use differentially rotating rings at the vertical
boundaries, or modify the boundary layer by localized Lorentz forces
\cite{BR70}.

A third issue is finite-amplitude or nonlinear hydrodynamical instability in
Rayleigh-stable regimes. Few theoretical studies on this subject exist 
\cite{SE59,JM70}.
It has been argued from experiments that a rapid Couette flow can be
nonlinearly unstable \cite{RZ99}. However, there are indications that
such instabilities
are caused by wall surface defects \cite{SG59}, which can be minimized.
In fact, it has been shown numerically that a positive angular-momentum
gradient strongly resists nonlinear instability
\cite{BH96,HB99}. The outcome depends, however,
on the amplitude of the initial perturbation and
the strength of the angular-momentum gradient.
These questions could be addressed empirically and relatively inexpensively
in prototype experiments using water.

One would like to predict the nonlinear phases of MRI in the
laboratory system. Ultimately, nonlinear MRI is closely related to other 
important physics of accretion disks involving magnetic field, 
i.e., dynamo processes and jet formation.
Ongoing three-dimensional MHD simulations \cite{KJ01} will provide useful insights here.
Indeed, an important purpose of the experiment
is to be a testbed for MHD codes since, lacking detailed observational
constraints, theorists depend upon computer simulations to understand
MRI-driven turbulence.

The experiment will be far more resistive
than most accretion disks, though perhaps not all \cite{GA96,GM98}.
Simulations indicate that when the field is
generated by the disk itself (magnetic dynamo),
then the large-scale field is nearly toroidal,
the important instabilities are nonaxisymmetric, and the turbulence 
sustains itself only if $S$ and $\Rm$ are much larger than
the experiment proposed here will achieve \cite{BH91b,BR95,HG96,SIM98,FS00}.
On the other hand, the innermost (and therefore most energetic)
parts of accretion disks often encounter a vertical field due to
the central compact object. The works cited above find that
in the presence of an imposed vertical field,
turbulence is driven by axisymmetric modes and
persists to higher resistivity, probably into the experimentally
accessible regime.

In summary, we have used linear stability analyses
to explore the prospects for
magnetorotational instability in a magnetized Couette flow.
We find that MRI can be achieved in a moderately rapidly rotating
table-top apparatus using an easy-to-handle liquid metal such as gallium.
Auxiliary experiments with an inexpensive
nonmagnetic fluid, such as water, will be valuable both as prototypes
and as controls to distinguish MRI from nonlinear hydrodynamic instabilities.
Onset and dynamics of MRI can be detected by torque measurements of cylinders
and magnetic sensors placed around the annulus.  Ultrasonic imaging
may also be possible.  If successful, this will be a rare
example of an astrophysical process that can be studied in the laboratory.

\section*{Acknowledgments}
The authors are grateful to Drs. P. Diamond, R. Goldston, S. Hsu, 
R. Kulsrud, W. Tang, and M. Yamada for fruitful discussions.
We thank our anonymous referee for the valuable comments, especially
those on the oscillatory modes.
This work is supported by U.S. Department of Energy
and by NASA grant NAG5-8385 [to JG].

{}

\begin{thebibliography}{}

\bibitem[\protect\citename{Balbus \& Hawley }1991a]{BH91} Balbus, S.A., Hawley, J.F. 1991, ApJ, 376, 214
\bibitem[\protect\citename{Balbus \& Hawley }1991b]{BH91b} Balbus, S.A., Hawley, J.F. 1991, ApJ, 376, 223
\bibitem[\protect\citename{Balbus \& Hawley }1998]{BH98} Balbus, S.A., Hawley, J.F. 1998, Rev. Mod. Phys., 70, 1
\bibitem[\protect\citename{Balbus, Hawley, \& Stone }1996]{BH96} 
	Balbus, S.A., Hawley, J.F., Stone, J.M. 1996, ApJ, 467, 76
\bibitem[\protect\citename{Blaes \& Balbus }1994]{BB94} Blaes, O.M., Balbus, S.A. 1994, ApJ, 421, 163
\bibitem[\protect\citename{Brandenburg et al.\ }1995]{BR95} 
	Brandenburg, A., Nordlund, \AA., Stein, R.F., Torkelsson, U. 1995, ApJ, 446, 741
\bibitem[\protect\citename{Brahme }1970]{BR70} Brahme, A. 1970, Physica Scripta, 2, 108
\bibitem[\protect\citename{Cabot }1996]{Ca96} Cabot, W. 1996, ApJ, 465, 874
\bibitem[\protect\citename{Chandrasekhar }1960]{CH60} Chandrasekhar, S. 1960, Proc. Nat. Acad. Sci., 46, 253
\bibitem[\protect\citename{Chandrasekhar }1961]{CH61} Chandrasekhar, S. 1961, Hydrodynamic and Hydromagnetic Stability,
	London: Oxford University Press
\bibitem[\protect\citename{Couette }1890]{CO90} Couette, T. 1890, Ann. Chim. Phys., 21, 433
\bibitem[\protect\citename{Curry, Pudritz, \& Sutherland }1994]{CP94} Curry, C., Pudritz, R.E., Sutherland, P.G. 
	1994, ApJ, 434, 206
\bibitem[\protect\citename{Donnelly \& Caldwell }1964]{DC64} Donnelly, R. J. Caldwell, D. R. 1964, 
	J. Fluid Mech., 19, 257
\bibitem[\protect\citename{Donnelly \& Ozima }1960]{DO60} Donnelly, R. J. Ozima, M. 1960, Phys. Rev. Lett., 4, 497
\bibitem[\protect\citename{Donnelly \& Ozima }1962]{DO62} Donnelly, R. J. Ozima, M. 1962, 
	Proc. R. Soc. Lond. A, 266, 272
\bibitem[\protect\citename{Fleming, Stone, \& Hawley }2000]{FS00} Fleming, T.P., Stone, J.M., Hawley, J.F. 
	2000, ApJ, 530, 464
\bibitem[\protect\citename{Gammie }1996]{GA96} Gammie, C.F. 1996, ApJ, 457, 355
\bibitem[\protect\citename{Gammie \& Menou }1998]{GM98} Gammie, C.F. 
\& Menou, K. 1998, ApJ, 492, L75
\bibitem[\protect\citename{Goodman \& Ji }2001]{GJ01} Goodman, J., Ji, H. 2001, submitted to J. Fluid Mech.
\bibitem[\protect\citename{Hartmann }1937]{HA37} Hartmann, J. 1937, Mat-Fys. Medd., 15, 6
\bibitem[\protect\citename{Hawley }2000]{HA00} Hawley, J.F. 2000, ApJ, 528, 462
\bibitem[\protect\citename{Hawley, Balbus, \& Winters }1999]{HB99} Hawley, J.F., Balbus, S.A., Winters, W.F. 1999, 
	ApJ, 518, 394
\bibitem[\protect\citename{Hawley, Gammie, \& Balbus }1996]{HG96} Hawley, J.F., Gammie, C.F., Balbus, S.A. 1996, 
	ApJ, 464, 690
\bibitem[\protect\citename{Jin }1996]{JI96} Jin, L. 1996, ApJ, 457, 798
\bibitem[\protect\citename{Joseph \& Munson }1970]{JM70} Joseph, D.D., Munson, B.R. 1970, J. Fluid Mech., 43, 545
\bibitem[\protect\citename{Kageyama, Ji, \& Goodman }2001]{KJ01} Kageyama, A., Ji, H., Goodman, J. 2001, 
	to be submitted
\bibitem[\protect\citename{Lynden-Bell \& Pringle }1974]{LP74} Lynden-Bell, D., Pringle, J.E. 1974, MNRAS, 168, 603
\bibitem[\protect\citename{Matsumoto \& Tajima }1995]{MT95} Matsumoto, R., Tajima, T. 1995, ApJ, 445, 767
\bibitem[\protect\citename{e.g. Meier, Koide, \& Uchida }2001]{MK01} Meier, D.L., Koide, S., Uchida, Y. 2001, 
	Sci, 291, 84
\bibitem[\protect\citename{Pringle }1981]{PR81} Pringle, J.E. 1981, ARA\&A, 19, 137
\bibitem[\protect\citename{Richard \& Zahn }1999]{RZ99} Richard, S., Zahn, J.P. 1999, A\&A, 347, 734
\bibitem[\protect\citename{Rayleigh }1916]{RA16} Rayleigh, L. 1916, Proc. R. Soc. Lond. A, 93, 148
\bibitem[\protect\citename{Sano, Inutsuku \& Miyama }1998]{SIM98} Sano, T., Inutsuku, S., \& Miyama, S. 1998, 
	ApJ, 506, L57
\bibitem[\protect\citename{Sano \& Miyama }1999]{SM99} Sano, T., Miyama, S. 1999, ApJ, 515, 776
\bibitem[\protect\citename{Schultz-Grunow }1959]{SG59} Schultz-Grunow, F. 1956, Z. angew. Math. Mech., 39, 101
\bibitem[\protect\citename{Serrin }1959]{SE59} Serrin, J. 1959, Arch. Ration. Mech. Anal., 3, 1
\bibitem[\protect\citename{Shakura \& Sunyaev }1973]{SS73} Shakura, N.I., Sunyaev, R.A. 1973, A\&A, 24, 337
\bibitem[\protect\citename{Stone et al.\ }1996]{SH96} 
	Stone, J.M., Hawley, J.F., Gammie, C.F., Balbus, S.A. 1996, ApJ, 463, 656
\bibitem[\protect\citename{Taylor }1923]{TA23} Taylor G.I. 1923, Phil. Trans. Roy. Soc. London A, 223, 289
\bibitem[\protect\citename{Velikhov }1959]{VE59} Velikhov, E.P. 1959, Sov. Phys. JETP 36, 995

\end{thebibliography}
\end{document}